\documentclass[12pt]{article}
\usepackage[sc]{mathpazo}
\usepackage[a4paper,top=2cm,bottom=2cm,left=2cm,right=2cm,includehead,includefoot]{geometry}
\usepackage{graphicx, color}
\usepackage{amsmath, amssymb, amsfonts}
\usepackage[compress]{natbib}
\usepackage{hyperref}
\usepackage{bm, bbm, dsfont}
\usepackage{tabularx}
\usepackage{caption}
\usepackage{pdfpages}
\usepackage{float}
\let\oldtabular\tabular 
\renewcommand{\tabular}{\footnotesize\oldtabular}

\newcommand{\E}{\text{E}}
\newcommand{\var}{\text{var}}

\newcommand{\cor}{\text{cor}}

\title{Pairwise likelihood estimation of latent autoregressive count models 
\\
\vspace{0.5cm}
\small{\textit{The final version of the paper has been published in Statistical Methods in Medical Research} \url{https://doi.org/10.1177/0962280220924068}}}
\author{
\begin{tabular}{ccc}
\large{Xanthi Pedeli\footnote{Athens University of Business and Economics, Athens, Greece}\;  and Cristiano Varin\footnote{Ca' Foscari University, Venice, Italy} }
\end{tabular}}
\date{\today}

\begin{document}

\maketitle

\abstract{
Latent autoregressive models are useful time series models for the analysis of infectious disease data. Evaluation of the likelihood function of latent autoregressive models is intractable and its approximation through simulation-based methods appears as a standard practice. Although simulation methods may make the inferential problem feasible, they are often computationally intensive and the quality of the numerical approximation may be difficult to assess. We consider instead a weighted pairwise likelihood approach and explore several computational and methodological aspects including 
estimation of robust standard errors and the role of numerical integration. 
The suggested approach is illustrated using monthly data on invasive meningococcal disease infection in Greece and Italy.}

\section{Introduction}\label{sect:intro}

The analysis of infectious disease data is essential for strengthening surveillance systems 
and supporting the response to public health threats. Statistical modelling of infectious disease data needs to account for a variety of aspects including seasonalities, trends, the effect of concomitant infectious diseases, human activities that may favour the disease diffusion and external factors such as weather conditions  \citep{lessler:2016}. Methods for the statistical analysis of infectious diseases have continuously enriched the literature since the early twentieth century \citep{kermack:1927}, see \citet{held:2013} for a detailed account.

Although the statistical literature emphasizes the need of proper accounting for serial correlation in time series of disease counts \citep{zeger:2006, unkel:2012}, regression models that assume independent observations are commonly used by national epidemiological surveillance systems (\emph{e.g.} \citet{noufaily:2013}). Failing to account for the presence of serial correlation may lead to incorrect inferential results. A popular illustration is the series of poliomyelitis infections in the United States of America during the period 1970--1983 \citep{zeger:1988}. Standard analysis of these data with a periodic Poisson regression model that includes annual and semi-annual seasonality indicates the presence of a strong statistically significant decreasing linear trend during the observation period. However, different models that extend periodic Poisson regression to consider serial correlation agree that the trend in poliomyelitis cases is not statistically significant \citep{davis:1999}. 

Among various approaches suggested in the literature to incorporate serial correlation in regression models for counts, this paper will focus on latent autoregressive models also known as parameter-driven models \citep{cox:1981} or state space models for counts \citep{durbin:2012}. These models are attractive because of their flexible hierarchical structure and the simple interpretation of the model parameters. In latent autoregressive models, the serial dependence is described through an unobserved stationary Gaussian autoregressive process. Inference in latent autoregressive models is cumbersome because the likelihood function is an intractable high-dimensional integral. Various simulation  strategies for approximate likelihood and Bayesian inference  have been suggested in the literature. Examples include the Monte Carlo expectation-maximization, Markov chain Monte Carlo simulation, penalized quasi-likelihood, importance sampling, Laplace and Gaussian approximations, see for instance \citet{davis:2016} and the references therein.

In this paper, we consider pairwise likelihood inference \citep{varin:2011} based on a combination of likelihoods for pairs of observations. The pairwise likelihood offers a significant reduction of the computational cost related to the ordinary likelihood by replacing the high-dimensional likelihood integral  with a limited set of double integrals. Pairwise likelihood methods have been already considered for fitting latent autoregressive models in \citet{varin:2009}, \citet{davis:2011} and \citet{ng:2011}. 

This paper aims at highlighting methodological and implementation aspects that play an important role to obtain nearly efficient and numerically stable fitting of latent autoregressive models for counts using the method of maximum pairwise likelihood.  The first issue considered in this paper is which pairs should be included in the pairwise likelihood for efficient estimation. We compare pairwise likelihoods that give the same weight to all the pairs until a fixed distance with pairwise likelihoods  that gradually downweight the contribution of pairs formed by more distant observations. 

The second issue regards the approximation of the asymptotic variance of maximum pairwise likelihood estimators.  We extend previous results for the maximum pairwise likelihood estimator built only on consecutive pairs described in \citet{davis:2011} to the general case of a pairwise likelihood that includes also non-consecutive pairs.

A third issue treated in this paper relates to the numerical evaluation of the pairwise likelihood. We employ Gauss-Hermite cubature for approximating the double integrals appearing in the pairwise likelihood function and discuss the effect of the number of quadrature nodes on the efficiency of the maximum pairwise likelihood estimators.

The rest of the paper is organized as follows. Section \ref{sect:latent-ar1} briefly reviews latent autoregressive models for counts and sets up the notation used through the paper. Section \ref{sect:pairfit} details the pairwise likelihood approach for the estimation of latent autoregressive models. Finite sample properties of the pairwise likelihood estimators are assessed with simulations in Section \ref{sect:simulations}. Section \ref{sect:application} illustrates the methodology with the analysis of the monthly counts of invasive meningococcal disease infections in the authors' countries, Greece and Italy. These time series were obtained from the Surveillance Atlas of the European Center of Disease Control  (\url{https://ecdc.europa.eu/en/surveillance-atlas-infectious-diseases}).  The methods discussed in this paper have been implemented in a \texttt{R} \citep{r:2016} package called \texttt{lacm} (`latent autoregressive count models') available at the CRAN repository (\url{https://cran.r-project.org/web/packages/lacm}). Supplementary materials include  additional simulation results and the code for replicating the analyses in Section \ref{sect:application} using our package \texttt{lacm}. 

\section{Latent autoregressive models for counts}\label{sect:latent-ar1}

Let $y_1, \ldots, y_n$ be an observed time series of length $n$. Latent autoregressive models are specified around an unobserved autoregressive Gaussian model 
\begin{equation}
\label{eq:state}
u_t =\phi u_{t-1} + \epsilon_t,
\end{equation}
with $\epsilon_t\sim N(0, \sigma^2)$ and $|\phi|<1$.  Conditionally on the unobserved $u_t$, the observed counts $y_t$ are assumed to be independent Poisson random variables  with conditional expectation 
\begin{equation}\label{eq:latentAR1}
\text{E}(y_t | u_t) = \exp({\bm x}_t ^{\text T} {\bm \beta} + u_t),
\end{equation}
where ${\bm x}_t$ is a vector of regressors and ${\bm \beta}=(\beta_0, \ldots, \beta_p)^{\text T}$ the corresponding vector of regression coefficients. The latent model (\ref{eq:latentAR1})  assumes that the linear predictor ${\bm x}_t ^{\text T} {\bm \beta}$ is imperfect and the role of the `error term' $u_t$ is to capture the missing information. The inclusion of the latent variable $u_t$ in the linear predictor has the double effect of inducing serial correlation  and overdispersion, a feature frequently observed in time series of disease counts (\emph{e.g.} \citet{imai:2015}). In fact, the marginal mean and variance of $Y_t$ are
\begin{equation}
\label{eq:mean-var}
\E(y_t)= \exp({\bm x}_t ^{\text T} {\bm \beta} + \tau^2 / 2) \quad \text{and} \quad
\var(y_t)= \E(y_t) + \E(y_t)^2 \{\exp(\tau^2)-1\} ,
\end{equation}
where $\tau^2=\sigma^2/(1-\phi^2)$ is the stationary variance of the latent process $u_t.$
Differently from linear models, the autocorrelation function
\begin{equation}\label{eq:corr}
\cor(y_t, y_{t-i})=\frac{\E(y_t)\E(y_{t-i})\{\exp(\phi^i\tau^2)-1\}}{\sqrt{\var(y_t) \var(y_{t-i})}}, \quad i \geq 1,
\end{equation}
depends on the marginal moments and therefore on the regressors.

Likelihood inference for latent autoregressive models requires to approximate the $n$-fold integral
\begin{equation}
\label{eq:full.lik}
L(\boldsymbol{\theta})=\int_{\mathbb{R}^n}p(y_1|u_1;\beta)p(u_1;\sigma^2,\phi)\prod_{t=2}^np(y_t|u_t;\beta)p(u_t|u_{t-1};\sigma^2,\phi)du_1\ldots du_n
\end{equation}
where 
${\bm \theta}=({\bm \beta}^\text{T}, \sigma^2, \phi)^\text{T}$ is the parameter vector. 

Alternatively, the likelihood can be expressed as a series of $n$ nested one-dimensional integrals using the so-called filtering algorithm described, for example, in \citet{cagnone:2017}. The likelihood  is expressed as the product of the predictive densities, that is
\begin{equation*}
\label{eq:predictive-lik}
L({\bm \theta})=p(y_1; {\bm \theta}) \prod_{t=2}^n p(y_t | y_{t-1}, \ldots, y_{1}; {\bm \theta}),
\end{equation*}
where 
$$
p(y_1; \theta)= \int_{\mathbb{R}} p(y_{1}|u_{1}; {\bm \beta}) p(u_1; \sigma^2, \phi) \, d u_1
$$
and 
\begin{equation}\label{eq:filter1}
p(y_t| y_{t-1}, \ldots, y_{1}; \theta)= \int_{\mathbb{R}}  p(y_{t}|u_{t}; {\bm \beta}) p(u_{t} | y_{t-1}, \ldots, y_{1}; {\bm \theta}) \, d u_t.
\end{equation}
The conditional density of $u_{t}$ given $(y_1, \ldots, y_{t-1})$ is computed as 
\begin{equation}\label{eq:filter2}
p(u_{t} | y_{t-1}, \ldots, y_{1}; {\bm \theta}) = \int_{\mathbb{R}} p(u_t | u_{t-1}; {\bm \theta}) p(u_{t-1} | y_{t-1}, \ldots, y_{1}; {\bm \theta})  \, d u_{t-1},
\end{equation}
with
\begin{equation}\label{eq:filter-update}
p(u_{t-1} | y_{t-1}, \ldots, y_{1}; {\bm \theta})  = \frac{p(y_{t-1}|u_{t-1}; {\bm \theta}) p(u_{t-1}|y_{t-2}, \ldots, y_{1}; {\bm \theta})}{p(y_{t-1}|y_{t-2}, \ldots, y_{1}; {\bm \theta})}.
\end{equation}
The filtering algorithm proceeds through recursive evaluation of integrals (\ref{eq:filter1}) and (\ref{eq:filter2}) using the updating formula (\ref{eq:filter-update}). The drawback of the filtering algorithm is the propagation of the numerical error through the nested integrals. This effect is implicitly observed in \citet{cagnone:2017}. They report that approximation of the integrals (\ref{eq:filter1}) and (\ref{eq:filter2}) by Gauss-Hermite integration is inexpensive but inaccurate and fails to converge when $\phi$ is moderate to large, that is when the propagation effect is more evident.  In contrary, the more expensive but also more accurate adaptive Gauss-Hermite integration works well for all values of $\phi.$

\section{Pairwise fitting of latent autoregressive models}\label{sect:pairfit}

\subsection{The pairwise likelihood of order d}\label{sect:pairlik}

The pairwise log-likelihood of order $d$\citep{varin:2009} is defined as the sum of the log bivariate densities for all the pairs of observations that are separated at most by $d$ units,  \begin{equation}
\label{eq:PL-linear}
\ell_d({\bm \theta})= \sum_{t=d+1}^{n} \sum_{i=1}^d \log p(y_{t-i},y_{t};{\bm \theta}),
\end{equation}
where 
\begin{equation}\label{eq:bivpair}
p(y_{t-i},y_{t};{\bm \theta}) =\int_{\mathbb{R}^2} p(y_{t}|u_{t}; {\bm \beta}) p(y_{t-i}|u_{t-i}; {\bm \beta}) p(u_{t-i}, u_t; \sigma^2, \phi)  d u_{t-i} d u_t.
\end{equation}
The pairwise likelihood of order $d$ offers a significant reduction of the computational cost related to the ordinary likelihood by replacing the high-dimensional integral (\ref{eq:full.lik}) with $(n-d)d$ double integrals. Illustrations of the performance of the pairwise likelihood of order $d$ in time series models are discussed in \citet{varin:2009}, \citet{davis:2011}, and \citet{ng:2011}.

\begin{figure}
\centering
\includegraphics[scale=0.8]{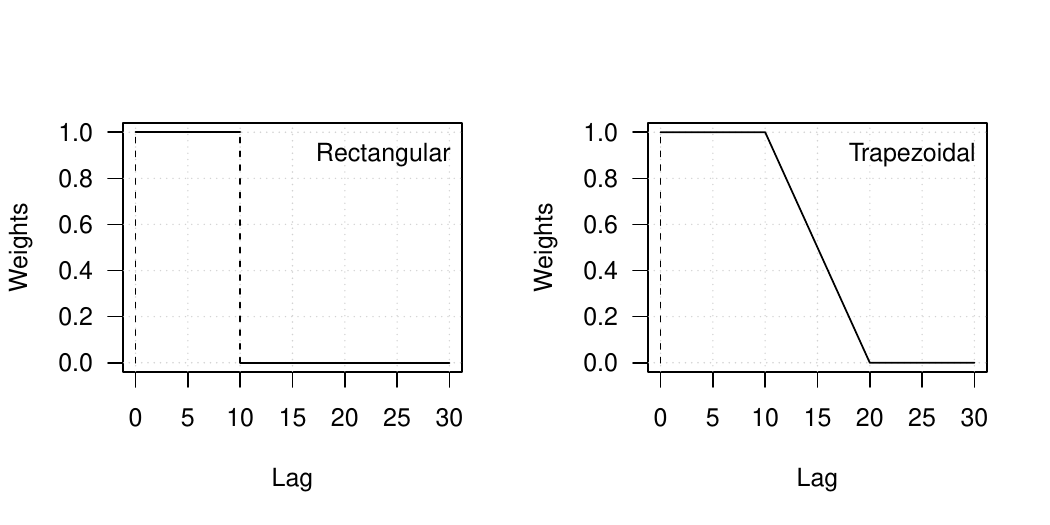} \\ 
\caption{Rectangular and trapezoidal (unnormalized) weights for the pairwise likelihood of order $d=10$.}\label{fig:trapezoidal}
\end{figure}

In this paper, the pairwise likelihood of order $d$ is rewritten as a weighted pairwise likelihood 
\begin{equation}
\label{eq:PL-linear}
\ell_d({\bm \theta})= \sum_{t=m_d+1}^{n} \sum_{i=1}^{m_d} w_i \log p(y_{t-i},y_{t};{\bm \theta}),
\end{equation}
for some non-negative weights $w_i$ and a window length parameter $m_d$ discussed below. We assume that the weights $w_i$ are normalized, so that $\sum_{i=1}^{m_d} w_i=1$. The pairwise likelihood of order $d$ corresponds to the rectangular weights
\begin{equation*}
w_i = \begin{cases}
1/d, & 1\leq i \leq d,\\
0, & \text{otherwise}.
\end{cases}
\end{equation*}
In this paper we also consider trapezoidal weights that linearly downweight the contribution from pairs of observations that are distant more than $d$ units. The trapezoidal weights are proportional to 
\begin{equation*} 
 w_i \propto 
 \begin{cases}
 1, & 1\leq i < d,\\
(2d-i)/d, & d\leq i < 2d,\\
 0, & i \geq 2d.
 \end{cases}
 \end{equation*}
The window length parameter $m_d$ is equal to $d$ for the rectangular weights and $2d$ for the trapezoidal weights. Figure \ref{fig:trapezoidal} illustrates rectangular and trapezoidal weights.
 
The maximum pairwise likelihood estimator of order $d$ is denoted as $\hat{{\bm \theta}}_{d}$ and it is the solution of  the pairwise score equations 
\begin{equation}\label{eq:scores}
\psi_d(\hat{\bm \theta}_d) = \sum_{t=m_d+1}^{n}  \psi_{d,t}(\hat{\bm \theta}_d)= {\bm 0},
\end{equation}
where $ \psi_{d,t}(\bm \theta)$ are the averaged pairwise scores  
\begin{equation*}
\psi_{d,t}(\bm \theta) = \sum_{i=1}^{m_d} w_i \frac{\partial }{\partial {\bm \theta}}\log p(y_{t-i},y_{t};{\bm \theta}).
\end{equation*}
As the order $d$ diverges, the pairwise likelihood involves an increasing number of pairs of independent observations that do not contain any information about the dependence parameter $\phi$ that therefore cannot be consistently estimated. Therefore, we consider $d$ fixed and study the limit behaviour of the pairwise likelihood estimator as $n$ diverges. 

The limit distribution of $\hat{{\bm \theta}}_{d}$ is derived through a direct extension of the results contained in \citet{davis:2011} for the pairwise likelihood of order one. The assumed regularity conditions  
are reported in \citet{ng:2011}
The Taylor series expansion of the pairwise score equations (\ref{eq:scores}) around the true value ${\bm \theta}_\ast$ gives
\begin{align*}
\sqrt{n} ( \hat{{\bm \theta}}_{d} - {\bm \theta}_\ast ) &=  \left\{ -\frac{1}{n}\frac{\partial}{\partial {\bm \theta}} \psi_d({\bm \theta}_\ast) \right\}^{-1} \frac{\psi_d({\bm \theta}_\ast)}{\sqrt{n}}\{1+ o_p(1)\} \\
&=\left\{ -\frac{1}{n} \sum_{t=m_d+1}^n \frac{\partial}{\partial {\bm \theta}} \psi_{d,t}({\bm \theta}_\ast) \right\}^{-1} \sqrt{n} \left\{ \frac{1}{n} \sum_{t=m_d+1}^n  \psi_{d, t}({\bm \theta}_\ast) \right\} \{1+ o_p(1)\}  .
\end{align*}
Since $\partial \psi_{d,t}({\bm \theta}_\ast) / \partial {\bm \theta}$ is an ergodic sequence, then
$$
-\frac{1}{n} \sum_{t=m_d+1}^n \frac{\partial}{\partial {\bm \theta}} \psi_{d,t}({\bm \theta}_\ast)= \E\left\{-\frac{\partial}{\partial {\bm \theta}}\psi_{d,t}({\bm \theta}_\ast)\right\} + o_p(1) ={\boldsymbol H}_d + o_p(1).
$$
Moreover, since the averaged pairwise scores $\psi_{d,t}({\bm \theta}_\ast)$ forms a stationary and strongly mixing sequence, then  $\psi_{d}({\bm \theta}_\ast)/ \sqrt{n}$ is asymptotically normal with zero mean and covariance matrix
$$
{\boldsymbol J}_d=\sum_{k=-\infty}^\infty \text{E}\left\{\psi_{d, t-k}({\bm \theta}_\ast) \psi_{d,t}({\bm \theta}_\ast)^\text{T} \right\}.
$$
Matrices ${\boldsymbol H}_d$ and ${\boldsymbol J}_d$ are referred as the sensitivity and variability matrices, respectively \citep{varin:2011}. 
The maximum pairwise likelihood estimator of order $d$ has therefore a limit normal distribution with mean ${\bm \theta}_\ast$ and asymptotic variance equal to the inverse of the Godambe information
$$
{\boldsymbol G}_d({\bm \theta}_\ast)={\boldsymbol H}_d({\bm \theta}_\ast) {\boldsymbol J}_d({\bm \theta}_\ast)^{-1} {\boldsymbol H}_d({\bm \theta}_\ast). 
$$
As noted by \citet{davis:2016}, the consistency and asymptotic normality of maximum pairwise likelihood estimation gives a potential advantage over standard maximum likelihood estimation whose limit properties for the latent autoregressive model have not yet been fully argued.

The sensitivity matrix ${\boldsymbol H}_d$ can be consistently estimated with the observed pairwise likelihood information
\begin{equation*}\label{eq:hessian}
\hat{\boldsymbol H}_d= -\frac{1}{n} \sum_{t=m_d+1}^n \frac{\partial}{\partial {\bm \theta}}\psi_{d,t}(\hat{{\bm \theta}}_d).
\end{equation*}
An alternative estimator of ${\boldsymbol H}_d$ that uses only first-order derivatives is derived from the second-order Bartlett identity that holds for each specific pair of observations and allows to rewrite the sensitivity matrix as
\begin{align*}
{\boldsymbol H}_d &= \E\left\{-\frac{\partial}{\partial {\bm \theta}}\psi_{d,t}({\bm \theta}_\ast)\right\}  \\
&= \sum_{i=1}^{m_d} w_{i} \E\left\{ -\frac{\partial^2}{\partial {\bm \theta} \partial{\bm \theta}^{\text T}} \log p(y_{t-i},y_{t};{\bm \theta}) \right\} \\
&= \sum_{i=1}^{m_d} w_{i}   \E\left\{  \frac{\partial}{\partial {\bm \theta}} \log p(y_{t-i},y_{t};{\bm \theta})  \frac{\partial}{\partial {\bm \theta}} \log p(y_{t-i},y_{t};{\bm \theta})^{\text T} \right\}.
\end{align*}
The above expression invites to consider the outer-product estimator 
\begin{equation}\label{eq:outer}
\hat{\boldsymbol H}_d= \frac{1}{n}\sum_{i=1}^{m_d}  w_i \left\{  \sum_{t=m_d+1}^n \frac{\partial}{\partial {\bm \theta}} \log p(y_{t-i},y_{t};\hat{{\bm \theta}}_d) 
 \frac{\partial}{\partial {\bm \theta}} \log p(y_{t-i},y_{t};\hat{{\bm \theta}}_d)^{\text T}
 \right\}.
\end{equation}
The outer-product estimator avoids the need of coding intricate second-order derivatives or to consider their computationally demanding and potentially unstable numerical approximations. Another advantage is that the outer-product estimator is guaranteed to be positive definite. 

Heteroskedasticity and autocorrelation consistent (HAC)  estimators developed in the econometric literature (\emph{e.g.}, \citet{neway:1994}) can be used for estimation of the variability matrix $\hat{\boldsymbol J}_d$ yielding
\begin{equation}
\label{eq:varmat}
\hat{\boldsymbol J}_d=\sum_{k=-r}^r \left(1-\frac{|k|}{r} \right) \left\{ \frac{1}{n} \sum_{t=m_d+1}^n \psi_{d, t-k}(\hat{{\bm \theta}}_d) \psi_{d,t}(\hat{{\bm \theta}}_d)^\text{T}  \right\}.
\end{equation}
The weights $(1-|k|/r)$ correspond to the popular Bartlett kernel. Other types of kernels like quadratic spectral, Parzen or truncated might also be used similarly to HAC estimators. Our experiments suggest that a reasonable choice is setting  the window semi-length $r$ equal to the default number of lags considered in correlograms produced by statistical softwares. Thereafter, we adopt the rule $r=\lfloor 10\log_{10}n\rfloor$  corresponding to the number of lags used in the \textbf{acf()} function of the \textbf{R} software\cite{r:2016}.

Estimators $\hat{\boldsymbol H}_d$ and $\hat{\boldsymbol J}_d$ are consistent outside model conditions, thus making the corresponding standard errors robust to model misspecification. 

Model selection can be performed with the composite likelihood information criterion \citep{varin:2011},  
$$
\text{CLIC} = - 2 \ell_d(\hat{\bm \theta}_d) + 2 \, \text{trace}(\hat{\boldsymbol H}_d^{-1} \hat{\boldsymbol J}_d). 
$$
Like the usual Akaike Information Criterion, models with lower values of CLIC are preferred.

\subsection{Numerical evaluation of the pairwise likelihood}\label{sect:cubature}

The simpler option for numerical evaluation of the bivariate integrals (\ref{eq:bivpair}) is the double Gauss-Hermite quadrature that has been used for computing pairwise likelihoods of latent autoregressive time series models in \citet{davis:2011}. Despite their popularity, the latent variable literature reports various difficulties with  Gauss-Hermite quadrature rules, see, for example,   \citet{bianconcini:2014}. \citet{rabe-hesketh:2005} point out that in random effects models with large cluster sizes or intraclass correlations, the poor performance of the Gauss-Hermite quadrature may be attributed to sharp peaks of the integrands located between adjacent quadrature points. \citet{lesaffre-spiessens:2001} note that in logistic random-intercepts models the validity of the method depends on the number of quadrature points.

A popular alternative to the standard Gauss-Hermite quadrature is its adaptive version that adjusts the quadrature locations using the mode and the curvature of the posterior density of the latent variables given the observations to provide a more accurate approximation of the integral \citep{liu:1994, bianconcini:2014}. 
Such an approach has been considered by \citet{ng:2011} for composite likelihood estimation of time series models with a latent autoregressive structure.
The application of adaptive Gauss-Hermite quadrature to pairwise likelihood requires to (i) maximize all the joint complete pairwise densities $p(y_t, y_{t-i}, u_t, u_{t-i}; {\bm \theta})$ with respect to the latent variables $(u_t, u_{t-i})$ and (ii) compute the Hessian matrix at the maximum for the various values of ${\bm \theta}$ explored by the optimization algorithm. Although adaptive Gauss-Hermite quadrature has proved to give more precise and stable results using a significantly smaller number of nodes than non-adaptive Gauss-Hermite quadrature, convergence to global maximum can be difficult to obtain even with relatively simple models \citep{lesaffre-spiessens:2001}. Moreover, 
evaluation of the benefits of adaptive Gauss-Hermite quadrature should also consider the cost of the possible large number of two-dimensional optimizations required to apply it to pairwise likelihood inference.

Another form of adaptive cubature known as h-adaptive is described in \citet{genz:1980} and \citet{berntsen:1991} and implemented in the \texttt{R} package \texttt{cubature} \citep{narasimhan:2017}. As for any adaptive integration method, 
the h-adaptive method recursively partitions the integration domain into smaller subdomains and the same integration rule is applied to each subdomain until convergence is achieved.
Subregion-adaptive integration methods are usually based on polynomial integrating basic rules that can provide rapid convergence once the subdivision is fine enough so that a low degree polynomial approximation can provide an accurate approximation to the integrand. 
Preconditioning the integrand by bringing the integration domain of the double integrals (\ref{eq:bivpair}) to the square unit makes the transformed integral easier to compute.

One important drawback of adaptive integration rules is that the resulting integral is discontinuous with respect to the model parameter ${\bm \theta}$ because the cubature nodes depend on ${\bm \theta}$ itself. The lack of continuity creates difficulties in the maximization of the approximated log-pairwise likelihood and makes unstable the evaluation of the  derivatives  in  the sensitivity and variability matrices ${\boldsymbol H}$  and  ${\bm J}$ needed for quantification of the estimation uncertainty and for model selection with the CLIC statistic. These difficulties, together with the computational cost of adaptive integration rules, hindered their use in the simulation experiments presented in the next section.

In contrast, the explicit calculation of first-order derivatives of the pairwise likelihood with the standard Gauss-Hermite quadrature stabilizes and speeds-up the computations in a way that the pairwise likelihood estimation turns out to be competitive to the popular integrated nested Laplace approximation (INLA)\citep{rue:2009} in terms of both the estimators' efficiency and computational time.

\section{Simulation studies}\label{sect:simulations}

We conducted a series of simulation experiments for assessing the finite sample properties of the maximum pairwise likelihood estimator. We considered the same simulation setting of \citet{davis:2005} and \citet{davis:2011}. Time series are simulated from a Poisson latent autoregressive model without covariates. The model parameters used in the simulations are characterized in terms of different levels of latent autocorrelation and different values of the dispersion index $D$ of the conditional variance of $y_t$ given $u_t$. The dispersion index is equal to $D_t=\E(y_t)(e^{\tau^2}-1)$ and summarizes the ability to extract information in the latent process $u_t$. Table \ref{tab:scenarios} lists the model parameters for  the nine simulation scenarios considered in \citet{davis:2005} and \citet{davis:2011}. Since no covariates are included in the model, the dispersion index is constant.  In the nine scenarios, the dispersion index is equal to 0.1, 1 or 10 and the latent autocorrelation parameter $\phi$ is -0.5, 0.5 or 0.9.

\begin{table}
\centering
\caption{Parameter values used in the simulation experiments. Each row corresponds to one of the nine scenarios considered in \citet{davis:2005}.}\label{tab:scenarios}
\begin{tabular}{r|r|rrr|rr}
  \hline
& $D$& $\beta$ & $\phi$ & $\sigma$ & $\tau^2$  \\ 
  \hline
1 & 10 & -0.6130 & -0.5000 & 1.2360 & 2.0369   \\ 
  2 & 10 & -0.6130 & 0.5000 & 1.2360 & 2.0369    \\ 
  3 & 10 & -0.6130 & 0.9000 & 0.6221 & 2.0369   \\ 
  \hline
  4 & 1 & 0.1501 & -0.5000 & 0.6190 & 0.5109   \\ 
  5 & 1 & 0.1501 & 0.5000 & 0.6190 & 0.5109   \\ 
  6 & 1 & 0.1501 & 0.9000 & 0.3115 & 0.5107   \\ 
  \hline
  7 & 0.1 & 0.3732 & -0.5000 & 0.2200 & 0.0645   \\ 
  8 & 0.1 & 0.3732 & 0.5000 & 0.2200 & 0.0645   \\ 
  9 & 0.1 & 0.3732 & 0.9000 & 0.1107 & 0.0645  \\ 
   \hline
\end{tabular}
\end{table}

For each of the nine scenarios, we generated 500  time series of size $n = 500$. Pairwise likelihoods are approximated using Gauss-Hermite quadrature with 5, 10, 20, 30 and 40 nodes per dimension. Maximization of the pairwise likelihood is performed with the BFGS algorithm with a relative convergence tolerance of $10^{-6}.$ Initial parameter estimates for the maximization of the pairwise likelihood are obtained using the method-of-moments estimator similarly to \citet{zeger:1988}; see also \citet{davis:1999} and \citet{ng:2011}. Maximum pairwise likelihood estimates were computed for orders $d$ from one to ten with both rectangular and trapezoidal weights. As a benchmark, we consider the integrated nested Laplace approximation \citep{rue:2009} as implemented in the \texttt{R} \citep{r:2016} package \texttt{INLA} (\url{www.r-inla.org}). 

\begin{figure}
\centering
\includegraphics[scale=0.8]{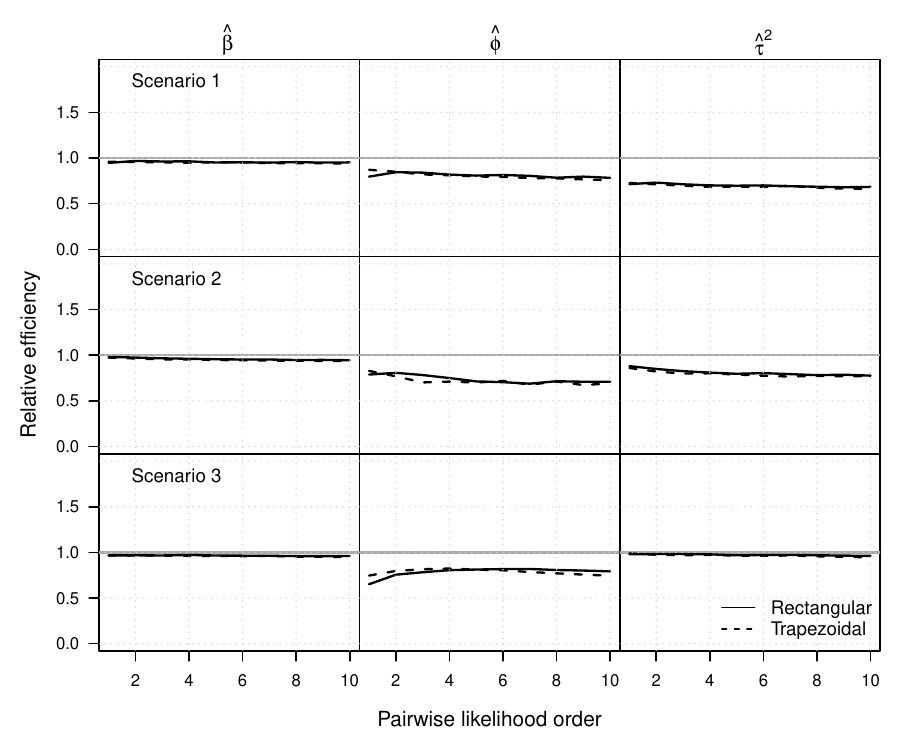} \\ 
\caption{Relative efficiency of maximum pairwise likelihood estimates with respect to INLA estimates as a function of the pairwise likelihood order $d$ for simulation scenarios 1 (upper panel), 2 (mid panel)  and 3 (bottom panel). The pairwise likelihood is approximated with Gauss-Hermite quadrature with 20 quadrature nodes per dimension using rectangular or trapezoidal weights.}\label{fig:sim123}
\end{figure}

\begin{figure}
\centering
\includegraphics[scale=0.8]{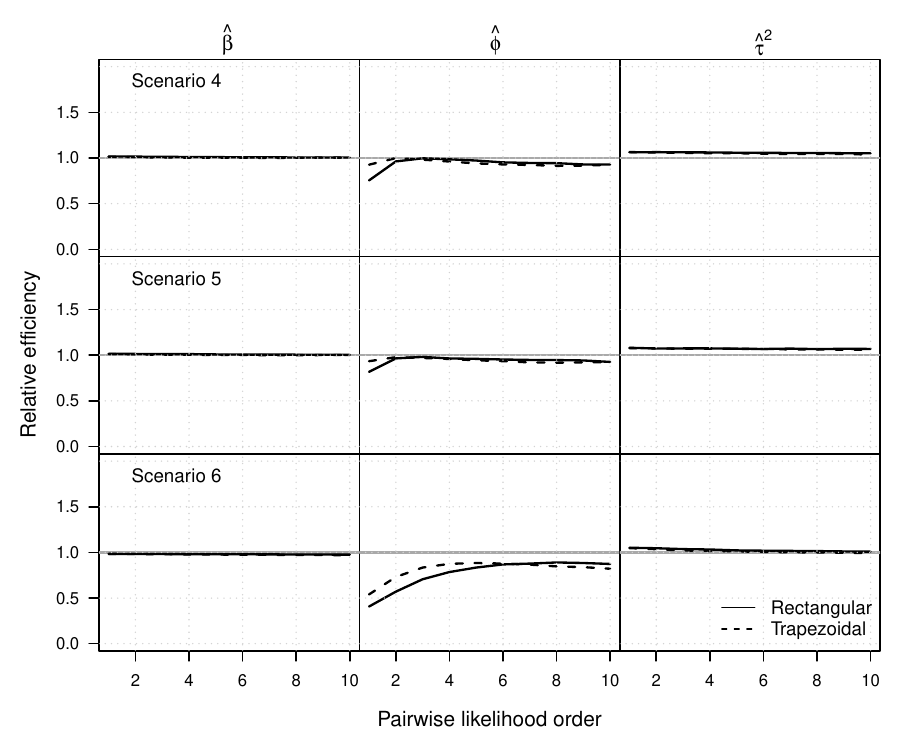} \\ 
\caption{Relative efficiency of maximum pairwise likelihood estimates with respect to INLA estimates as a function of the pairwise likelihood order $d$ for simulation scenarios 4 (upper panel), 5 (mid panel)  and 6 (bottom panel). The pairwise likelihood is approximated with Gauss-Hermite quadrature with 20 quadrature nodes per dimension using rectangular or trapezoidal weights.}\label{fig:sim456}
\end{figure}

\begin{figure}
\centering
\includegraphics[scale=0.8]{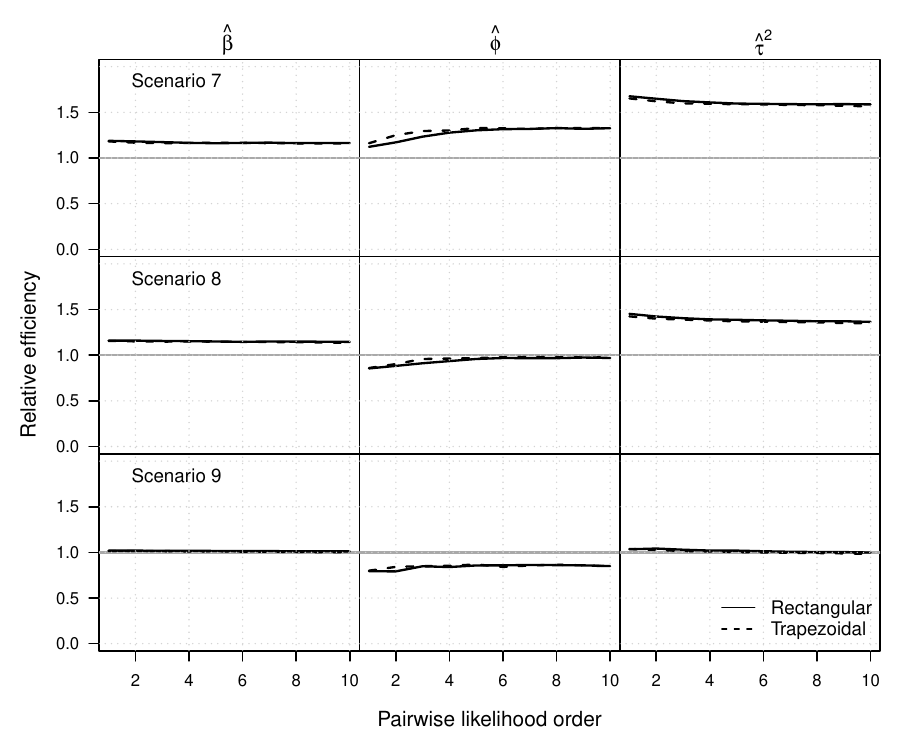} \\ 
\caption{Relative efficiency of maximum pairwise likelihood estimates with respect to INLA estimates as a function of the pairwise likelihood order $d$ for simulation scenarios 7 (upper panel), 8 (mid panel)  and 9 (bottom panel). The pairwise likelihood is approximated with Gauss-Hermite quadrature with 20 quadrature nodes per dimension using rectangular or trapezoidal weights.}\label{fig:sim789}
\end{figure}

Figures \ref{fig:sim123},  \ref{fig:sim456} and \ref{fig:sim789} display the relative efficiency of maximum pairwise likelihood estimates as a function of the pairwise likelihood order, when the pairwise likelihood is approximated with 20 quadrature nodes per dimension. The relative efficiency is measured with the ratio of the root mean square error of maximum pairwise likelihood to the root mean square error of INLA estimates. Values of the ratio above one indicate that maximum pairwise likelihood performs better than INLA in terms of mean square error.

According to the results illustrated in Figures \ref{fig:sim123}--\ref{fig:sim789}, the dispersion index  $D$ plays a decisive role in the efficiency of the maximum pairwise likelihood estimators. In particular, for $D=10$ (Figure \ref{fig:sim123}), the INLA estimates of $\phi$ and $\tau^2$ perform better than the corresponding maximum pairwise likelihood estimates, while the efficiency of $\hat{\beta}$ is almost the same with the two estimation approaches. For $D=1$ (Figure \ref{fig:sim456}), the maximum pairwise likelihood and INLA estimates are almost equivalent, especially when $\phi$ equals 0.5 in absolute value. For $D=0.1$ (Figure \ref{fig:sim789}) the maximum pairwise likelihood estimates of $\beta$ and $\tau^2$ outperform the corresponding INLA estimates. Regarding the parameter $\hat{\phi}$, the maximum pairwise likelihood approach achieves a higher efficiency when $\phi=0.5$, almost equal efficiency to INLA when $\phi=-0.5$ and lower efficiency when $\phi=0.9$. The overall conclusion is that maximum pairwise likelihood estimation is competitive with INLA for estimation of autoregressive latent models for counts. 

Our results indicate that the order $d$ of the pairwise likelihood is  irrelevant for estimation of $\beta$, it has a small effect on estimation of $\tau^2$ and it is quite important for estimation of $\phi$. More precisely, the relative efficiency of $\hat{\tau}^2$ tends to slightly reduce with the order, while the relative efficiency  of $\hat{\phi}$ increases until some order and afterwards stabilize or even decrease.

 Figures \ref{fig:sim123}--\ref{fig:sim789} also allows for a comparison between rectangular and trapezoidal weights. The linear downweighting of the contributions from pairs of observations that are distant more than $d$ units gives some improvement on the relative efficiency of $\hat{\phi}$, especially when $D=1$ (Figure \ref{fig:sim456}) and $\phi=0.9$. Estimation of $\beta$ and $\tau^2$ seems to be insensitive to the weights.


Figures SM1-SM4 in the Supplementary Materials reveal another interesting role of the dispersion index. These figures regard the first three scenarios of Table \ref{tab:scenarios} where $D=10$ and display the relative efficiency of the maximum pairwise likelihood estimates  when the pairwise likelihood is approximated with 5, 10, 30 and 40 quadrature nodes per dimension. A comparative inspection of Figures SM1-SM4 together with Figure \ref{fig:sim123} -- that reports the simulation results with 20 nodes per dimension -- suggests that the relative efficiency of maximum pairwise likelihood estimation increases with the number of quadrature nodes. The simulations also suggest that using less than 20 nodes per dimension leads to an inadequate approximation of the pairwise likelihood resulting in numerical instabilities in the corresponding maximum pairwise likelihood estimates.

However, the case is different when $D=0.1$ or $D=1$. The simulation results indicate that just a few quadrature nodes are sufficient when the dispersion index is small. In particular, the root mean square errors of the simulated maximum pairwise likelihood estimates are unchanged up to  three digits for 5 to 40 quadrature nodes per dimension. Thus, in this case, the pairwise likelihood can be adequately approximated by Gauss-Hermite quadrature with even 5 nodes per dimension which implies a significant reduction in the already short time required for maximum pairwise likelihood fitting. Since the plots of the relative root mean square errors as a function of the quadrature nodes  are essentially  indistinguishable when $D=0.1$ and $D=1$, then they are not reported in the Supplementary Materials. 

Although the simulations suggest that a limited number of quadrature nodes can be safely used only in settings with a small dispersion index, the applicability of this result is fairly broad because our own experience is that in epidemiological time series the dispersion index is rarely large otherwise the regression component will be uninformative, as illustrated in the next section.

Finally, we observe that our finding about the need to increase the number of quadrature nodes when the dispersion index is large is in line with the results in \citet{lesaffre-spiessens:2001}.

\section{Applications}\label{sect:application}

\subsection{Invasive meningococcal disease in Greece}\label{sect:greece}

Invasive meningococcus infection is a rare  but severe disease with relatively high case fatality and up to one fifth of all survivors suffering from long-term sequelae \citep{rosenstein:2001}. The surveillance of  invasive meningococcal disease in Europe is coordinated by the European Center of Disease Control (ECDC) with input of case-based data reports from national surveillance systems.
Thereafter, we illustrate latent autoregressive models in surveillance using data on the monthly number of meningococcal disease cases in Greece for the years 1999-2016. The data source is the ECDC Surveillance Atlas. We employ the time series until 2015 for model fitting and then compute the predictions of invasive meningitis cases in 2016.

\begin{figure}
\centering
\includegraphics[scale=0.8]{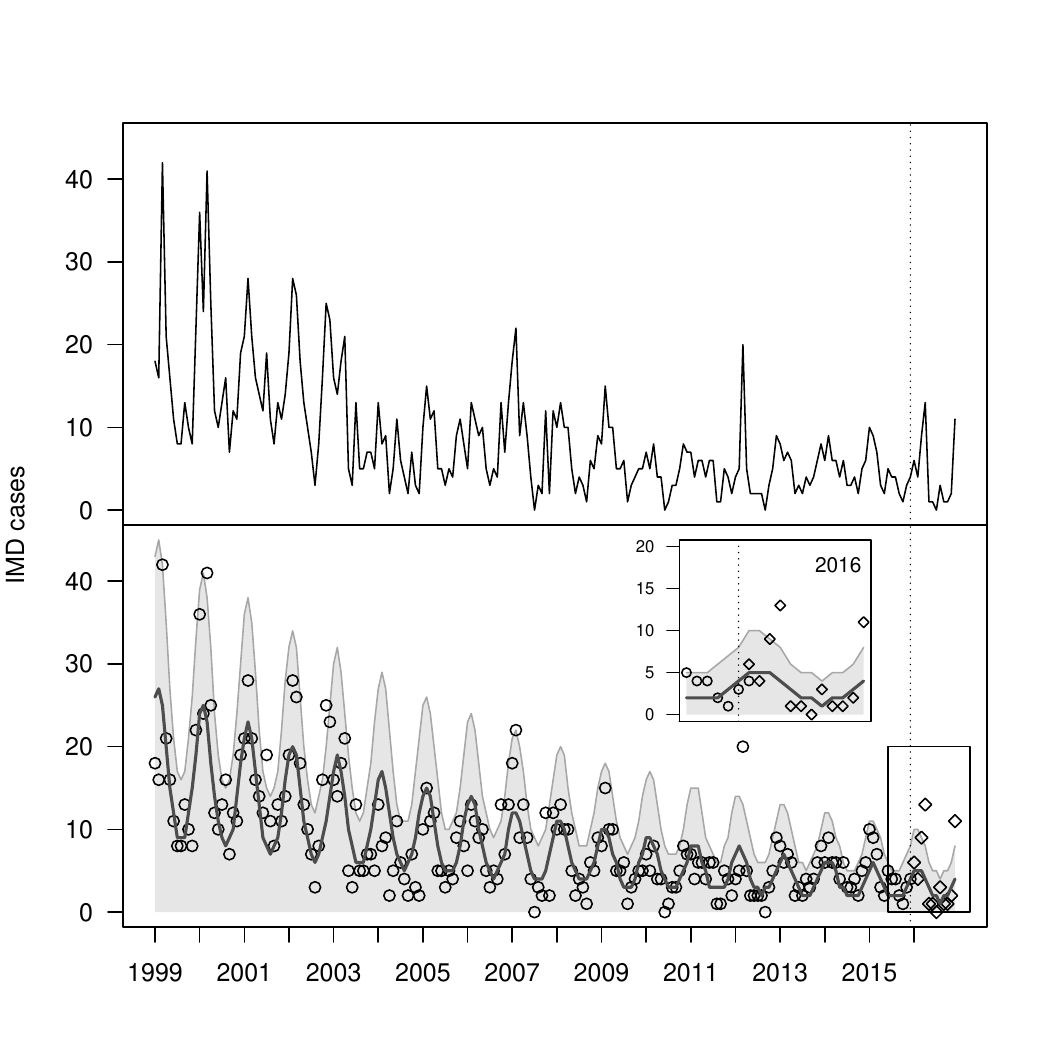}
\caption{Upper panel: Time series of the monthly number of invasive meningococcal disease (IMD) cases in Greece for the years 1999--2016. Lower panel: observed ($\circ$) and predicted (--) number of IMD cases until year 2015. The observed cases in 2016  are denoted with symbol $\diamond$. The shaded area corresponds to the $95\%$ upper tail prediction interval. The vertical dotted line separates the data used for model fitting from the data used for the prediction exercise. The inset plot in the lower panel is a magnification of the out-of-sample predictions for year 2016.}
\label{fig:meningo-gr}
\end{figure}

The plot of the monthly number of cases of invasive meningococcal disease in Greece in the upper panel of Figure \ref{fig:meningo-gr} reveals annual seasonality and a decreasing time trend. 

We model the Greek invasive meningococcal disease time series with a Poisson regression model with conditional expectation $\E(y_t | u_t)=\exp(\eta_t+u_t)$, where the linear predictor $\eta_t$  accounts for the annual seasonality and trend,
\begin{equation*}
\label{eq:linearpredgr}
\eta_t=\beta_0+\beta_1\frac{t}{204}+\beta_2\sin\left(\frac{2\pi t}{12}\right)+\beta_3\cos\left(\frac{2\pi t}{12}\right).
\end{equation*}
The trend term has been scaled by the number of observations (204 months) in way to report the model coefficients approximatively on the same scale. 

\begin{figure}
\centering
\includegraphics[scale=0.55]{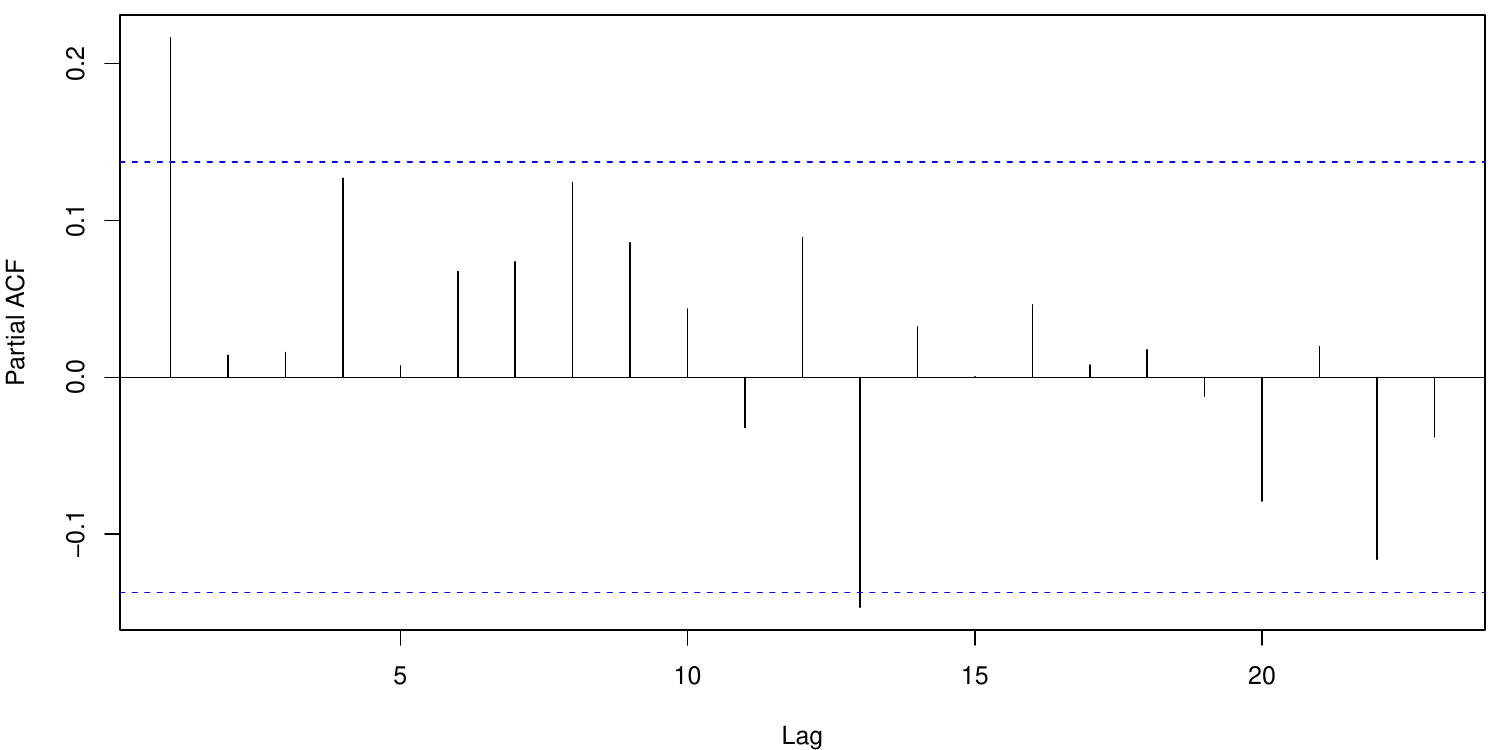} 
\caption{Partial correlogram of the residuals of  the independence model fitted on the series of of the monthly number of invasive meningococcal disease (IMD) cases in Greece for the years 1999--2016.}
\label{fig:pacf-gr}
\end{figure}

The order $d$ of the pairwise likelhood should be chosen in way to include only those pairs of observations that are genuinely informative, avoiding the pairs that are spuriously correlated.  Figure \ref{fig:pacf-gr} displays the sample partial autocorrelation function of the Pearson residuals of the standard Poisson regression model with linear predictor $\eta_t$. The plot indicates the presence of non-spurious correlation between pairs of observations separated by one lag. Accordingly, we fit the latent autoregressive model with the pairwise likelihood of order one. Given the results of our simulation studies, we adopt the trapezoidal weights.   

Numerical integration for computation of the pairwise likelihood is performed through Gauss-Hermite quadrature with 5, 10 and 20 nodes per dimension giving the same estimates and standard errors up to two decimal digits. This result is coherent with our findings in the simulation studies because a preliminary estimation of the dispersion indices $D_t$ using the method-of-moments gives values that range from 0.14 to 2.13 with a median value of 0.54.  These values of $D_t$ indicate that a low number of quadrature nodes is sufficient for approximation of the bivariate integrals of the pairwise likelihood.

The parameter estimates and the corresponding standard errors obtained with  the latent autoregressive model fitted with the pairwise likelihood of order one with 10 nodes per dimension are displayed in Table \ref{tab:summary-gr}. For comparison purposes, we report also the results obtained with INLA.   

\begin{table}
\caption{Parameter estimates (standard errors) for models fitted to the monthly counts of meningococcal infections in Greece for the period 1999-2015. The estimation methods are maximum pairwise likelihood (MPLE) or INLA.}
\label{tab:summary-gr}
\centering
\begin{tabular}{lcc}
  \hline
& MPLE ($d=1$) &INLA\\
\hline
intercept & 2.79 (0.11) &2.75 (0.10)\\
trend &   -1.67 (0.22)  & -1.60 (0.18)\\
sin term & 0.46 (0.04) & 0.46 (0.05)\\
cosine term & 0.28 (0.04) & 0.27 (0.05)\\
$\phi$  & 0.52 (0.17) & 0.60 (0.14)\\
$\tau^2$ & 0.07 (0.02) & 0.07 (0.02)\\
\hline
\end{tabular}
\end{table}

As expected, the results confirm significant seasonality and a decreasing trend of the invasive meningococcal disease cases in Greece. Maximum pairwise likelihood estimates and standard errors with Gauss-Hermite quadrature are in close agreement with results based on INLA. 

Comparison between different fitting methods in terms of computational time requires care because it depends on various implementation aspects. In particular, the comparison between the computational times of our specialised algorithm for latent autoregressive models and INLA  is quite unbalanced because INLA is designed to efficiently fit a much wider class of models. However, the CPU time for maximum pairwise likelihood estimation on a 2017 MacBook Pro 13 with a 3.5 GhZ Intel Core i7 dual-core processor and 16 GB 2133 MHz memory is 0.047 CPU seconds with five quadrature nodes per dimension, 0.153 CPU seconds with ten quadrature nodes per dimension and 0.524 CPU seconds with twenty quadrature nodes per dimension compared to 2.490 CPU seconds needed for estimation with INLA.

The CLIC statistic can be used to evaluate the relevance of the latent model component. The CLIC statistic for the latent autoregressive model is 2117.4, a value somehow smaller than the CLIC for the reduced model without autocorrelation obtained setting $\phi=0$ (CLIC=$2121.6$) and much smaller than the CLIC for the independence model obtained by dropping the latent model component,   corresponding to $\tau^2=0$ and $\phi=0$ (CLIC=$2176.3$).

The lower panel of Figure \ref{fig:meningo-gr} displays the observed and predicted meningococcal disease cases in Greece together with the corresponding $95\%$ upper prediction bound computed with 10,000 simulations from the model fitted by maximum pairwise likelihood. The comparison of observations and in-sample predictions until the year 2015 in the lower panel of Figure \ref{fig:meningo-gr} indicates a realistic model fitting and retrospectively identifies a few outbreaks. We also report in the lower panel of Figure \ref{fig:meningo-gr} the predicted meningococcal disease cases and corresponding prediction upper bounds for the twelve months in 2016.  Out-of-sample predictions are close to the observed meningitis cases for the year 2016 except for April and December where the observed cases marginally exceed the corresponding 95\% upper prediction limit.

\subsection{Invasive meningococcal disease in Italy}\label{sect:italy}

\begin{figure}
\centering
\includegraphics[scale=0.8]{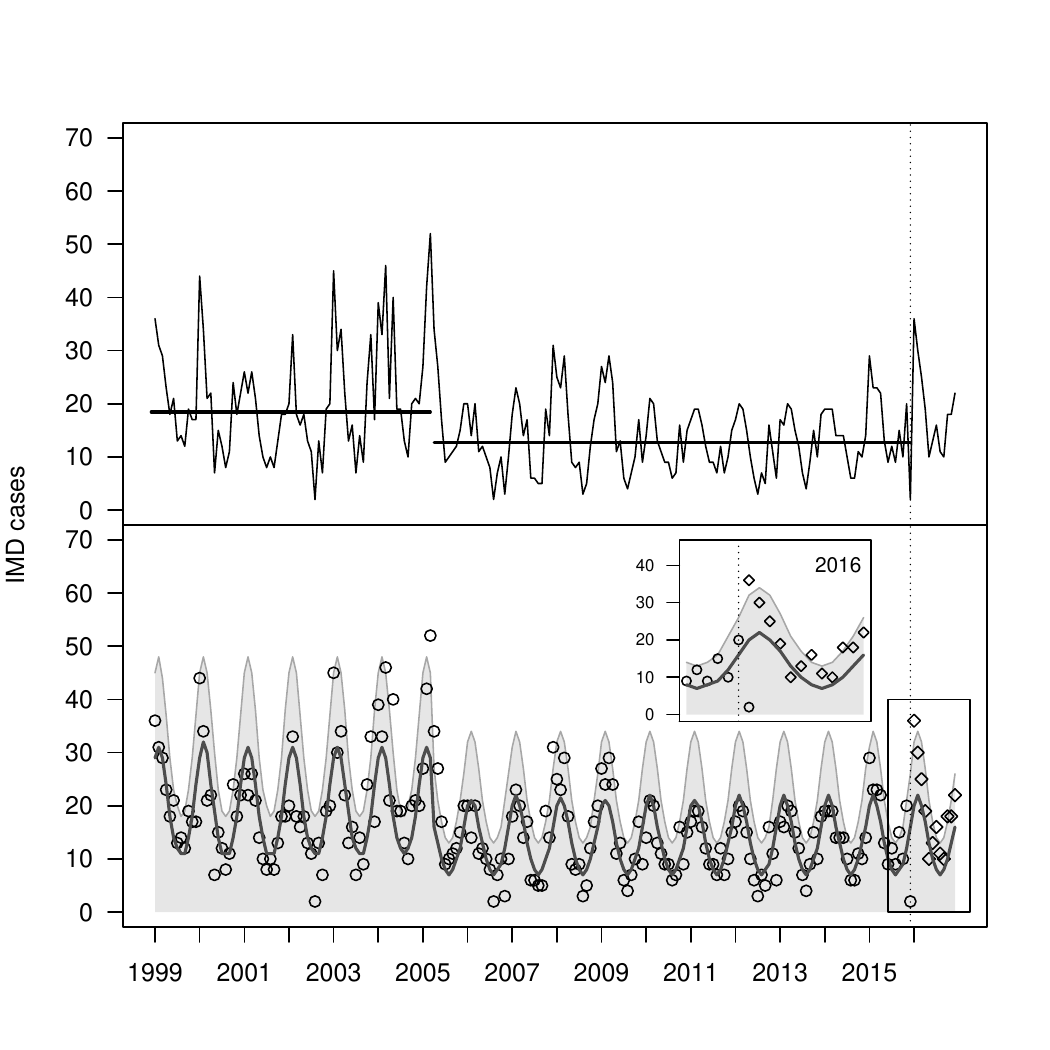}
\caption{Upper panel: Time series of the monthly number of invasive meningococcal disease (IMD) cases in Italy for the years 1999--2016. The horizontal lines indicate the average number of cases before and after March 2005. Lower panel: observed ($\circ$) and predicted (--) number of IMD cases until year 2015. The observed cases in 2016  are denoted with symbol $\diamond$. The shaded area corresponds to the $95\%$ upper tail prediction interval. The vertical dotted line separates the data used for model fitting from the data used for the prediction exercise. The inset plot in the lower panel is a magnification of the out-of-sample predictions for year 2016.}
\label{fig:meningo-it}
\end{figure}

The upper panel of Figure \ref{fig:meningo-it} shows the time series plot of the monthly number of meningococcal disease cases in Italy for the years 1999-2016.

The main feature of the Italian series is a level shift corresponding to a sensible reduction of the monthly number of cases after March 2005. This reduction can be explained as a consequence of increasing vaccination coverage due to the 2005--2007 National Italian Vaccine Plan in which the conjugate MenC vaccination was recommended to Italian citizens \citep{stefanelli:2009, pascucci:2014}. 

We consider the latent autoregressive model $E(y_t|u_t)=\exp(\eta_t+u_t)$ with linear predictor
\begin{equation*}
\label{eq:linearpredit}
\eta_t=\beta_0+\beta_1 x_t+\beta_2\sin\left(\frac{2\pi t}{12}\right)+\beta_3\cos\left(\frac{2\pi t}{12}\right),
\end{equation*}
where $x_t$ is a binary indicator that distinguishes between observations before ($x_t=1$)  and after ($x_t=0$)  March 2005. 

\begin{figure}
\centering
\includegraphics[scale=0.55]{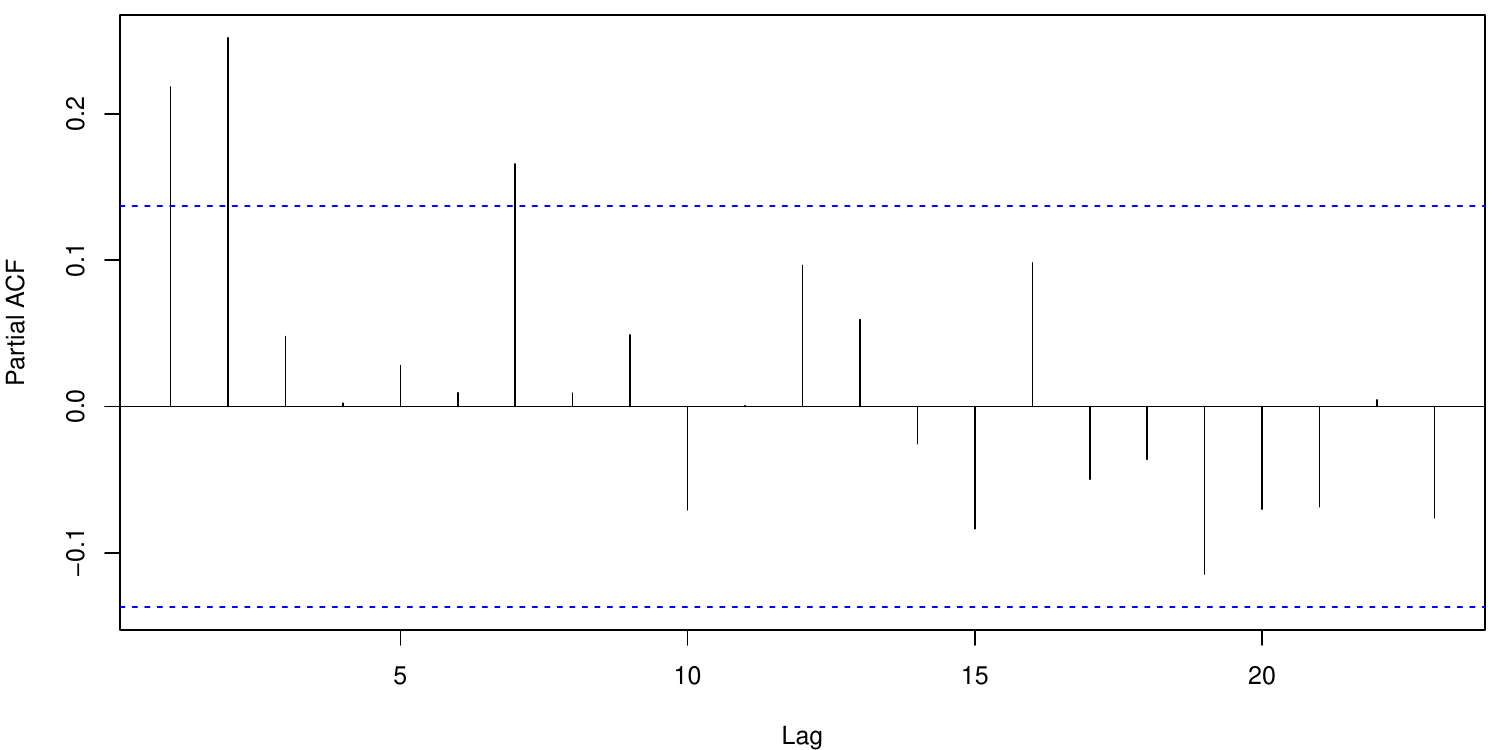} 
\caption{Partial correlogram of the residuals of  the independence model fitted on the series of of the monthly number of invasive meningococcal disease (IMD) cases in Italy for the years 1999--2016.}
\label{fig:pacf-it}
\end{figure}

Figure  \ref{fig:pacf-it} displays the sample partial autocorrelation function of the Pearson residuals of the standard Poisson regression model with linear predictor $\eta_t$. The sample partial autocorrelation indicates the presence of non-spurious correlation at the first two lags. Accordingly, we fit the latent autoregressive model with the pairwise likelihood of order two using trapezoidal weights. 

As for the Greek series, the preliminary estimates of the dispersion indices $D_t$ suggest that a low number of quadrature nodes is sufficient for approximation of the bivariate integrals involved in the pairwise likelihood: the minimum, maximum and median of the preliminary estimates of $D_t$ are equal to 0.38, 1.62 and 0.73 respectively. In fact, estimates and standard errors computed with 5, 10 and 20 quadrature nodes per dimension coincide up to two significant digits. Table \ref{tab:summary-it} summarizes the parameter estimates and standard errors obtained by maximization of the pairwise likelihood of order two with 10 nodes per dimension. Again maximum pairwise likelihood estimates are in close agreement with INLA and confirm the significant level shift in the invasive meningitis cases after March 2015. 

The CPU times for obtaining the maximum pairwise likelihood estimates and the corresponding standard errors -- using the same notebook employed for the analysis of the Greek series -- were 0.072, 0.235 and 0.896 with  five, ten and twenty quadrature nodes per dimension, respectively. Estimation with INLA required  2.443 CPU seconds. 

The CLIC statistics support the latent autoregressive model (CLIC$= 2423.1$) compared to the model without autocorrelation (CLIC$= 2430.4$) and the standard generalized linear model (CLIC$= 2510.0$).

\begin{table}
\centering
\caption{Parameter estimates (standard errors) for models fitted to the monthly counts of meningococcal infections in Italy for the period 1999-2015. The estimation methods are maximum pairwise likelihood (MPLE) or INLA.}
\label{tab:summary-it}
\centering
\begin{tabular}{lcc}
  \hline
&MPLE ($d=2$) &INLA\\
\hline
intercept & 2.54 (0.04) & 2.55 (0.06)\\
level shift & 0.38 (0.07) & 0.36 (0.09)\\
sin term & 0.47 (0.03) & 0.46 (0.04)\\
cosine term & 0.26 (0.04) & 0.26 (0.04)\\
$\phi$ &  0.70 (0.08) & 0.74 (0.11)\\
$\tau^2$ & 0.04 (0.01) & 0.04 (0.01)\\
\hline
\end{tabular}
\end{table}

The observed and predicted cases of meningococcal infections in Italy together with the corresponding $95\%$ upper bound intervals are illustrated in the lower panel of Figure \ref{fig:meningo-it}. Like for the Greek data, predictions were computed with 10,000 simulations from the fitted model. The in-sample predictions show that the latent autoregressive model fits well the observed data. 
The comparison of the out-of-sample predictions with the observed disease counts for the year 2016  does not formally indicate any outbreak since there are no observations that exceed the corresponding 95\% upper prediction limit. However, most the observations are above the prediction line and often close to the upper prediction limit. This non-random pattern is related to the unusual large number of invasive meningococcal disease cases   diagnosed in the Tuscany region between 2015 and 2016 \citep{stefanelli:2016}.

\section{Conclusions}
This paper discussed several methodological and implementation aspects of pairwise likelihood inference for latent autoregressive models for counts. We considered the general case of a pairwise likelihood of order $d$ and studied the efficiency of the maximum pairwise likelihood estimates. Empirical results indicated that maximum pairwise likelihood estimation is competitive to the popular integrated nested Laplace approximation regarding both the efficiency of the estimators and the computational time.

We also got some rough evidence that the use of weights that downweight the contribution of pairs of observations that are far apart in time could improve the efficiency of maximum pairwise likelihood estimation of the autoregressive parameter. Results in spatial and spatio-temporal settings \citep{bevilacqua:2012} suggest that non-rectangular weights  could be  useful in case of more complicated dependence structures, for example when the latent process is a Gaussian autoregressive process of order $p>1$. In this case, the use of appropriate weighting schemes together with sparsity constraints might improve the efficiency of maximum pairwise likelihood estimators while preserving a competitive computational cost. 

For the approximation of the double integrals involved in the pairwise likelihood function we adopted standard Gauss-Hermite quadrature. We found that the quality of the pairwise likelihood approximation is affected by the number of quadrature nodes only when the dispersion index of the conditional variance of $y_t$ given $u_t$ is large. However, we expect the dispersion index to be relatively small in typical applications of latent autoregressive count models so that in practice a limited number of quadrature nodes is sufficient in most of the cases.

This paper assumed that the latent process is a simple autoregressive model of order one, as most of the papers about latent count time series modelling. Robust standard errors were considered to protect inference on the regression parameters from the risk of misspecification of the dependence structure. While the extension of weighted pairwise likelihood inference to latent autoregressive models of order $p$ is straightforward, see for example \citet{ng:2011}, more difficult seems to be the case of a latent process with a moving average component. In fact, previous studies \citep{varin:2009, davis:2011} found that maximum pairwise likelihood estimation for moving average processes performs poorly with a considerable loss of efficiency. 

As the amount of quality surveillance datasets increases, there is also a growing interest
towards joint modelling of multiple diseases in spatio-temporal settings \citep{baker:2017}. The reduction
of the computational cost obtained with composite likelihood methods is particularly
attractive for fitting spatio-temporal extensions of the latent autoregressive model for
multiple diseases. In such contexts, composite likelihoods formed by blocks of data might be considered to describe complex dependence structures. Gauss-Hermite quadrature may not be appropriate for approximation of blockwise composite likelihood because it suffers from the curse of dimensionality. One interesting option is to consider sparse grids integration rules \citep{heiss:2008} that have attracted considerable attention in the econometric literature due to their generality, ease of implementation and low computational cost.


 \section*{Supplementary Materials}
Supplementary Materials contain additional figures related to the simulation study of Section \ref{sect:simulations} and the code for replicating the analyses in Section \ref{sect:application}.

\section*{Acknowledgements}
This project has received funding from the European Union’s Horizon 2020 research
and innovation programme under the Marie Sk{\l}odowska--Curie grant agreement
no.~699980 and the Athens University of Economics \& Business, Action I Funding.


  \newpage
\includepdf[pages=-]{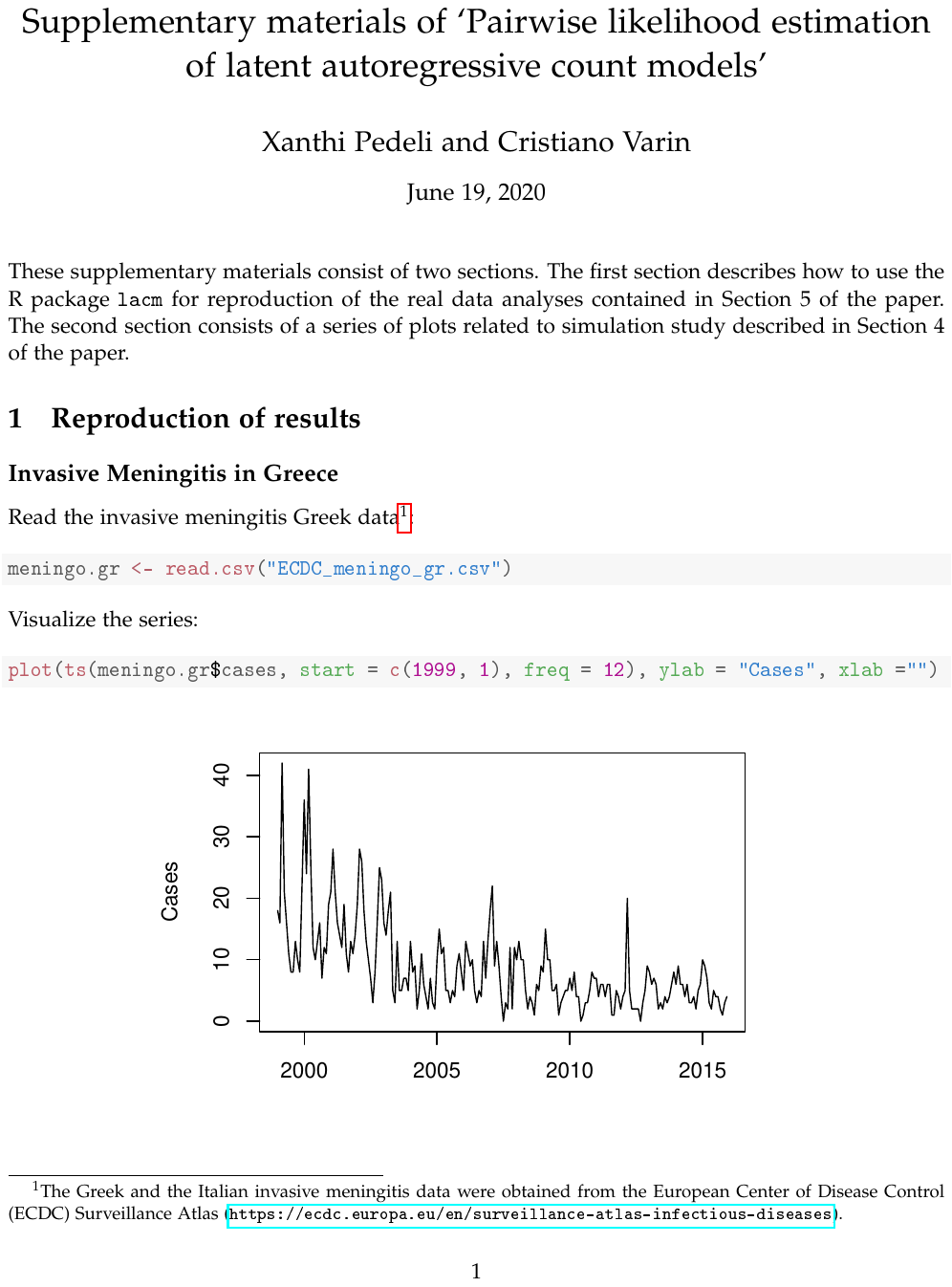}

\end{document}